\documentclass[12pt,preprint]{aastex}

\usepackage{graphicx}
\usepackage{txfonts}

\newcommand{\hi}{\ion{H}{1}}
\newcommand{\nai}{\ion{Na}{1}}

\newcommand{\mgi}{\ion{Mg}{1}}
\newcommand{\fei}{\ion{Fe}{1}}
\newcommand{\feii}{\ion{Fe}{2}}
\newcommand{\sii}{[\ion{S}{2}]}
\newcommand{\hei}{\ion{He}{1}}
\newcommand{\caii}{\ion{Ca}{2}}
\newcommand{\oi}{[\ion{O}{1}]}

\newcommand{\av}{$A_V$}

\newcommand{\msun}{M$_{\odot}$}
\newcommand{\rsun}{R$_{\odot}$}
\newcommand{\msunyr}{M$_{\odot}$\,yr$^{-1}$}
\newcommand{\macc}{$\dot{M}_{acc}$}

\newcommand{\mloss}{$\dot{M}_{loss}$}

\newcommand{\lacc}{$L_{\mathrm{acc}}$}
\newcommand{\rstar}{$R_{\mathrm{*}}$}
\newcommand{\rin}{$R_{_\mathrm{in}}$}

\newcommand{\mstar}{$M_{\mathrm{*}}$}

\slugcomment{To appear in Ap. J.}

\begin{document}

\title{On the 2015 outburst of the EXor variable star V1118 Ori}

\author{T.Giannini\altaffilmark{1}, 
D.Lorenzetti\altaffilmark{1},
S.Antoniucci\altaffilmark{1},
A.A.Arkharov\altaffilmark{2},
V.M.Larionov\altaffilmark{2,3},
A.Di Paola\altaffilmark{1},
S.Bisogni\altaffilmark{4},
and
A.Marchetti\altaffilmark{5}
}
\altaffiltext{1}{INAF - Osservatorio Astronomico di Roma, via Frascati 33, I-00078 Monte Porzio, Italy}
%
\altaffiltext{2}{Central Astronomical Observatory of Pulkovo, Pulkovskoe shosse 65, 196140 St.Petersburg, Russia}
\altaffiltext{3}{Astronomical Institute of St.Petersburg University, Russia}
\altaffiltext{4}{INAF - Osservatorio Astrofisico di Arcetri, Largo E. Fermi 5, I-50125 Firenze, Italy}

\altaffiltext{5}{INAF - Osservatorio Astronomico di Brera, Via Brera 28, I-20122 Milano, Italy}


\begin{abstract}
After a long-lasting period of quiescence of about a decade, the source V1118 Ori, one of the most representative members of the EXor variables, is now outbursting. Since the initial increase of the near-infrared flux of about 1 mag ($JHK$ bands) registered on 2015 September 22, the source brightness has remained fairly stable. We estimate $\Delta$V $\sim$ 3 mag with respect to the quiescence phase. An optical/near-IR low-resolution spectrum has been obtained with the Large Binocular Telescope instruments MODS and LUCI2, and 
compared with a spectrum of similar spectral resolution and sensitivity level taken during quiescence. Together with the enhancement of the continuum, the outburst spectrum presents a definitely higher number of emission lines, in particular \hi\, recombination lines of the Balmer, Paschen, and Brackett series, along with bright permitted lines of several species, forbidden atomic lines, and CO ro-vibrational lines.
Both mass accretion and mass loss rates have significantly increased (by to about an order of magnitude: \macc\,=\,1.2-4.8 10$^{-8}$\, \msunyr, \mloss\,= 0.8$-$2 10$^{-9}$\,\msunyr) with respect to the quiescence phase. 
If compared with previous outbursts, the present one appears less energetic. Alternatively, it could already be in the fading phase (with the maximum brightness level reached when the source was not visible), or, viceversa, still in the rising phase.
\end{abstract}

\keywords{stars: pre-main sequence --- stars: variables: T Tauri, Herbig Ae/Be --- stars: formation
--- accretion, accretion disks --- infrared: stars --- stars: individual (V1118 Ori)}

\section{Introduction}

Last phases of matter accretion of low-to-intermediate mass (0.5-8 $M_{\odot}$) young stellar objects (YSOs) are characterized by magnetospheric accretion events (Shu et al. 1994). Optical and near-IR observations have shown that the accretion process proceeds through irregular and rapid
outbursts, with amplitudes in the range 0.2$-$1 magnitudes.  
However, a small percentage of sources, namely the EXor variables (Herbig 1989), undergo much stronger and repetitive
outbursts imputed to a sudden increase of the mass accretion rate (e.g. Hartmann \& Kenyon 1985, Antoniucci et al. 2008). The typical brightness variation is of $\sim$ 4$-$5 magnitudes in the visual band, with a duration from few months to years and a recurrence time of several years. During the outburst phase EXor spectra are characterized by emission line spectra, from which accretion rates of    
$\sim$ 10$^{-6}-10^{-8}$ \msunyr\, are typically estimated (e.g. Lorenzetti et al. 2009; Sicilia-Aguilar et al. 2012; K\'{o}sp\'{a}l et al. 2011; Audard et al. 2014).
Such timescale variability makes EXors suitable candidates to perform studies based on evolutionary more than statistical concepts, aimed at answering the following fundamental questions: (1) which is the mechanism(s) triggering the intermittent accretion bursts?; (2) how does the mass accretion process eventually halt?; (3) what is the role of bursts in the evolution of the circumstellar disk, especially regarding the formation of proto-planetary systems? 

The poor characterization of the EXor class derives not only from the small number of known objects (around 20$-$30 objects, e.g. Lorenzetti et al. 2012; Audard et al 2014), but also from the lack of a long-term monitoring of the photometric and spectroscopic features of the sources. For these reasons we have started the program EXORCISM (EXOR optiCal and Infrared Systematic Monitoring,  Antoniucci et al. 2014), which is based on the 0.3-2.5 $\mu$m photometric and spectroscopic monitoring of about 20 objects, classified as EXors or recognized as suitable candidates.
Given the cadence of EXor outbursts, quiescence vs. outburst phases have been rarely observed with
instrumentation having the same performances: as a consequence, the results derived for both states are not meaningfully comparable. 

This is not the case of V1118 Ori, an EXor whose long-lasting quiescence phase has been recently investigated (Lorenzetti et al. 2015a, hereafter Paper I) by using an instrumentation similar to that employed  for
monitoring the outburst presented here. During the last four decades V1118 Ori underwent five outbursts, whose average duration was about a couple of years (1982-84, 1988-90, 1992-94, 1997-98, 2004-06). The first four eruptions are reported by Parsamian et al. 1993, Garc\'{i}a Garc\'{i}a \& Parsamian 2000; Herbig 2008 and references therein, whereas the properties shown during the fifth event are illustrated by Audard et al. (2005, 2010) and Lorenzetti et al. (2006, 2007). Here we account for the last outburst, which was announced few months ago in a telegram by Lorenzetti et al. (2015b). We describe in Section 2 our optical and near-IR observations;  while in Section 3 we analyze and discuss the results. Our concluding remarks are given in Section 4.

\section{Observations and results}

\subsection{Near-IR photometry} 

\begin{figure}
\epsscale{0.9}
\plotone{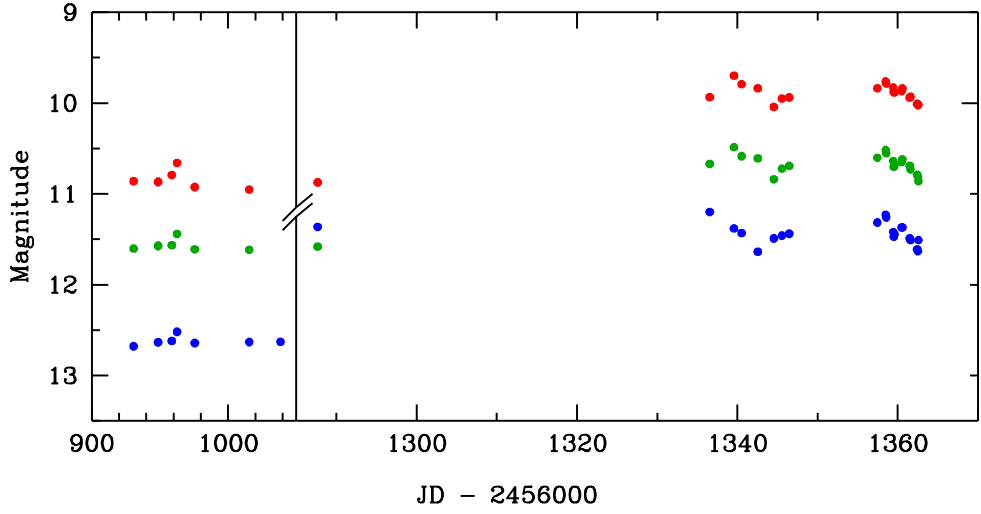}
\caption{V1118 Ori $JHK$ lightcurve (blue, green and red, respectively).\label{fig:fig1}}
\end{figure}

Near-IR data were obtained with the 1.1m AZT-24 telescope at Campo Imperatore  (L'Aquila, Italy) equipped with the
imager/spectrometer SWIRCAM (D'Alessio et al. 2000), which is based on a 256$\times$256 HgCdTe PICNIC array. $JHK$ photometry 
was obtained in the period September-December 2015. Reduction was achieved  by using standard procedures for bad pixel cleaning, flat fielding, and sky subtraction. The results are reported in Table~\ref{tab:tab1} and shown in Figure\,\ref{fig:fig1}. Averaged differences in magnitudes with respect to the last phase of quiescence (September-October 2014, Paper I) are $\Delta$J=1.23, $\Delta$H=0.92, $\Delta$K=0.99.

\subsection{Optical spectroscopy}

We obtained an optical spectrum with the 8.4m Large Binocular Telescope (LBT) using the Multi-Object Double Spectrograph (MODS - Pogge et al. 2010) on 2015 October 1st and 2nd (JD 2\,457\,296/97). Each of the two observations was performed with the dual grating mode (Blue + Red channels, spectral range 0.35$-$0.95 $\mu$m) and integrated 1500 s by using a 0$\farcs$8 slit ($\Re \sim$ 1500). 
Each spectral image was independently bias and flat-field corrected, then the sky background was subtracted in the two-dimensional spectral images. The final spectrum was obtained as the average of the two one-dimensional spectra, each obtained by collapsing the stellar trace in the 2-D image along the spatial direction. 
Wavelength calibration was obtained from the spectra of arc lamps, while flux calibration was achieved from observations of spectro-photometric standard stars.

\subsection{Near-IR spectroscopy}
A near-IR spectrum of V1118 Ori was obtained with the LUCI2 instrument at LBT on 2015 October 28th (JD 2\,457\,323).
The observations were carried out with the G200 low-resolution grating coupled with the 1$\farcs$0 slit. Two datasets were acquired with the standard ABB'A' technique using the $zJ$ and $HK$ filters, for a total integration time of 12 and 8 minutes, respectively. This provides a final spectrum covering the wavelength range 1.0$-$2.4 $\mu$m at $\Re \sim$1000.
The raw spectral images were flat-fielded, sky-subtracted, and corrected for optical distortions
in both the spatial and spectral directions. Telluric absorptions were removed using the normalized spectrum of a telluric standard star, after removal of its intrinsic spectral features.
Wavelength calibration was obtained from arc lamps, while flux calibration was based on the photometric data taken
in the same span of days. Both optical and near-IR outburst spectra are plotted in Figure\,\ref{fig:fig2} together with those obtained during the quiescence phase (taken from Paper I).

\section{Analysis and discussion}

\subsection{Photometry}
The light-curve shown in Figure\,\ref{fig:fig1} indicates that after a long period of quiescence that lasted about 10 years, V1118 Ori is currently in a high-level state. Our observations evidence the beginning of the new outburst between January and September 2015; in the subsequent three months the source has remained at an approximately constant level. The average near-IR color variation with respect to the quiescence state is $\Delta$[$J-H$] = 0.3 mag, $\Delta$[$H-K$] = $-$0.06 mag, which is roughly orthogonal to the extinction vector (see Figure\,4 of Paper I). As observed in a number of other classical EXors, this 
implies that extinction variations, although likely present,  do not play a major role during the outburst (see e.g. Figure\,1 of Lorenzetti et al. 2012). 
Indeed, also in the occasion of the 2005 outburst, when the near-IR colors were very similar to the present ones, Audard et al. (2010) did not measure significant extinction variations from the quiescence to the outburst phase. By means of the X-rays column density, they estimated A$_V$ $\sim$ 1$-$2 mag, which are assumed also by us in the present work. 
A rough estimate of the optical magnitudes has been retrieved from the MODS spectrum.  The continuum level increases by a factor between $\sim$ 40 and 4 passing from $U$ to $I$ band. Approximate $UBVRI$ magnitudes in outburst  (quiescence) are $U$\,=\,15.8 (19.8), $B$\,=\,15.8 (19.1), $V$\,=\,15.0 (18.0), $R$\,=\,14.2 (16.9), $I$\,=\,13.3 (14.8).  Given that the seeing during the nights of the observations never exceeded  1\farcs1, no significant flux losses have occurred, so the uncertainties on the estimated magnitudes are below 0.10-0.15 mag.
The [$U-B$], [$B-V$] colors change from 0.7 and 1.1 mag in quiescence to 0.0 and 0.8 mag in outburst.  In previous outbursts, however, the optical colors have become even bluer, being, for example, [$U-B$]=$-$0.84 and [$U-B$]=$-$0.94 in 1983 and in 1989 (Parsamian et al. 2002).
This fact indicates that the present outburst is one of the least energetic among those observed in V1118 Ori. Indeed, the visual magnitudes registered at the peak of previous outburst have been:  $V$\,=\,13.8 (1983, Marsden 1984); $V$\,=\,12.8 (1988, Gasparian \& Ohanian 1989); $V$\,$\le$\,14.7  (1992-1994,  Garc\'{i}a Garc\'{i}a et al. 1995); $V$=\,13.5  (1997, Hayakawa et al. 1998); $V\sim$\,12.8  (2005, Audard et al. 2010). Considering however that both the rising and declining times observed in the 1998 and 2005 outbursts have lasted $\sim$ 100$-$300 days (Garc\'{i}a Garc\'{i}a \& Parsamian 2000, Audard et al. 2010), it may be that the maximum level of activity was reached between January and September 2015 (when no observations are available), or that the outburst is still in the rising phase.  Very recent observations by Audard et al. (2016) indicate that no significant photometric variations in both optical and near-infrared bands have occurred until mid-January 2016.

\subsection{Outburst vs. quiescence spectrum}
\begin{figure}
\epsscale{1.0}
\plotone{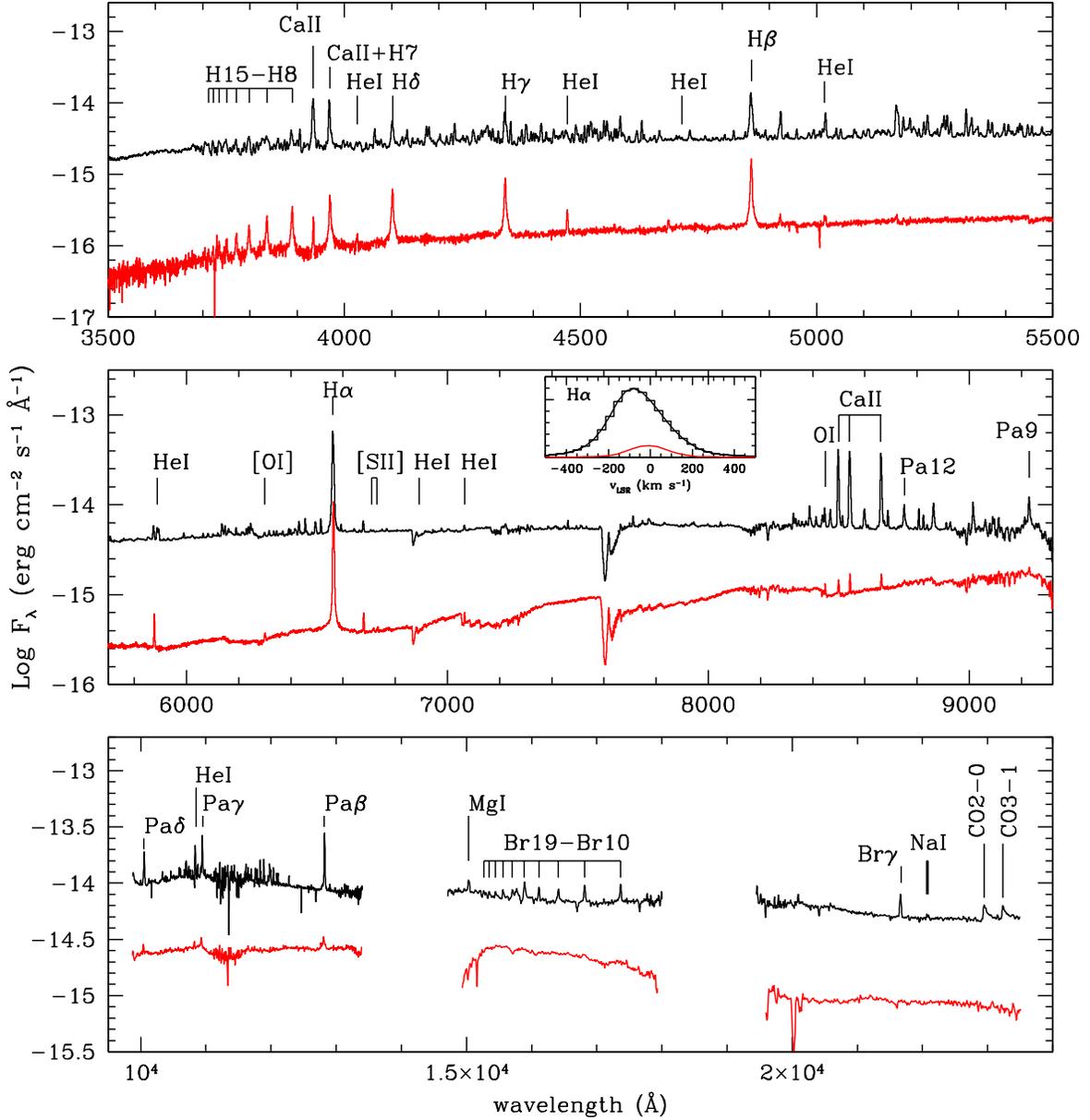}
\caption{Optical (LBT/MODS) and near-IR (LBT/LUCI2) spectrum of V1118 Ori in the present outburst phase (black) shown in comparison with the quiescence spectrum (red, spectra obtained with LBT/MODS and TNG/NICS, see Paper I). Main emission lines are labeled. Regions corrupted by atmospheric absorption were removed. The inset in the middle panel  shows the H$\alpha$ spectral profile. \label{fig:fig2}}
\end{figure}

The analysis of the optical/near-IR outburst spectrum was essentially done in comparison with the quiescence spectrum discussed in Paper I (shown in Figure~\ref{fig:fig2}). This was obtained with a similar instrumental setup as far as spectral resolution and sensitivity level are concerned. Notably, these are one of the few examples of quiescence and outburst spectra of the same source taken with the same high-level instrumentation.

A remarkable number of differences are evidenced by the comparison of the two spectra.
(1) A  much increased number of emission lines is recognizable in the outburst spectrum, mainly of neutral and ionized metals (\hi, \hei, \caii,
 \fei, \feii). Main lines are listed in Table\,\ref{tab:tab2} and their fluxes and equivalent widths (EW) are compared with those measured in the 
 quiescence spectrum (columns 5 and 6).  The outburst-to-quiescence flux ratio of the Balmer lines decreases from $\sim$ 30 to $\sim$ 8 going 
 from $n_{\rm{up}}$\,=\,15 to $n_{\rm{up}}$\,=\,3 and that of the Paschen series  from $\sim$ 30 to $\sim$ 12 going from $n_{\rm{up}}$\,=\,12 to 
 $n_{\rm{up}}$\,=\,7. Although a detailed modeling of the hydrogen line ratios is far from the aims of this paper, a qualitative comparison with 
 models of Paschen (Edwards et al. 2013) and Balmer (Antoniucci et al. 2016) decrement series suggests that the observed  behavior may be explained 
 by a 0.5$-$0.6 dex increase of the electron density, with no evident dependence on temperature variations. The ratio of the outburst-to-quiescence EW of the Balmer series lines 
 does not depend on n$_{\rm{up}}$ and is less than unity, suggesting that the line flux increase tightly follows that of the continuum. Conversely,
the EW ratios of the Paschen lines are between 2 and 6, indicating that  the increase of the  continuum level is smaller (and therefore slower) in the
infrared than in the optical. The \caii\, lines show the largest variations, with  $F_{\rm{out}}$/$F_{\rm{quiesc}}$ $\sim$ 85 and $EW_{\rm{out}}$/$EW_{\rm{quiesc}} \sim$ 15. (2) Forbidden lines of \oi\, and \sii, which remained  un-detected (or barely detected) in the quiescence spectrum are present in the outburst optical spectrum,  suggesting that an enhancement of the mass ejection activity might have been triggered by the increase of the mass accretion rate. (3) A weak Balmer jump (BJ), which is considered a direct indicator of accretion (e.g. Manara et al. 2016 and references therein) is spotted at the end of the Balmer series.
We measured the BJ by taking the ratio of the flux at 360 nm to that at 400 nm.  We obtained a value of 0.9, more than double the quiescence value. Also, this exceeds the limit of 0.5 empirically put by
 Herczeg \& Hillenbrand (2008) to identify mid-M dwarfs accretors. (4) In the near-IR part of the spectrum,  \mgi\, and \nai\, in emission, as well as CO overtone emission ($v$\,=\,2-0 and $v$\,=\,3-1), are clearly detected, while they were not present in the quiescence spectrum.
(5) Despite the moderate spectral resolution, we are able to detect differences in the spectral profile of the brightest \hi\, lines. As an example we show in Figure\,\ref{fig:fig2} the H$\alpha$ spectral profile. In the quiescence spectrum, the line is 
unresolved, and the emission peak is close to the rest velocity. During outburst, the line peak is blue-shifted by $\approx$ $-$70 km s$^{-1}$ (with respect to the local standard of rest), with a  FWHM $\sim$ 320 km s$^{-1}$, indicating the appearance of a fast wind in this phase. A similar behavior was exhibited by the H$\alpha$ profile during the 2005 outburst and subsequent decay (Herbig  2008). 
\\

\subsection{Comparison with spectra of previous outbursts}

The 2005 outburst was spectroscopically investigated in the optical and in the near-IR by Herbig (2008) and Lorenzetti et al. (2006), respectively. 
The optical spectrum was taken by Herbig during the declining phase (2005 November), when
the star was somewhat brighter ($V$ = 14.5) than at the epoch of our observation. Since the EWs of the most prominent lines (\hi\,, \hei\,, and \caii\,, see Herbig's 2008 table 3) were slightly lower than the present ones, we qualitatively deduce that the line fluxes are rather similar in the two outbursts. Herbig (2008) reports the unusual detection of Li\,I\,$\lambda$6707 {\it in emission}, a line commonly seen in absorption in classical T Tauri stars. We signal the presence of this line also in the MODS spectrum, although detected at a low signal-to-noise level  (Flux\,=\,1.5$\pm$0.4 10$^{-15}$ erg s$^{-1}$ cm$^{-2}$).\\

The near-IR spectrum was taken in September 2005 ($J$\,=\,11.2, Lorenzetti et al. 2006). The same emission features, with similar line fluxes as in the present case, were detected. In particular, since CO emission is believed to originate in the gaseous inner disk where it traces relatively warm ($\sim$4000 K) zones at high densities (Scoville et al. 1980), the constancy of the emitted flux could signal that the physical
conditions of the environment around the accretion channels remains similar during all the events, and only the accretion rate varies. Finally, in the same 2005 spectrum the H$_2$ 1-0S(1)  was barely detected. This line appears at around 1.5\,$\sigma$ level also in the LUCI2 spectrum. The line, together with the  H$_2$ 2-1S(1) line,  however, is definitely detected in the 2-D spectral image in a region extending from the source itself up to a distance of $\sim$ 17\arcsec. The flux ratio between the two lines in such region is $\sim$ 2, suggesting  a fluorescent excitation (e.g. Black \& van Dishoeck 1987) arising in the nebular environment where the source is located.

\subsection{Mass accretion and mass loss rates}
\begin{figure}
\epsscale{1.0}
\plotone{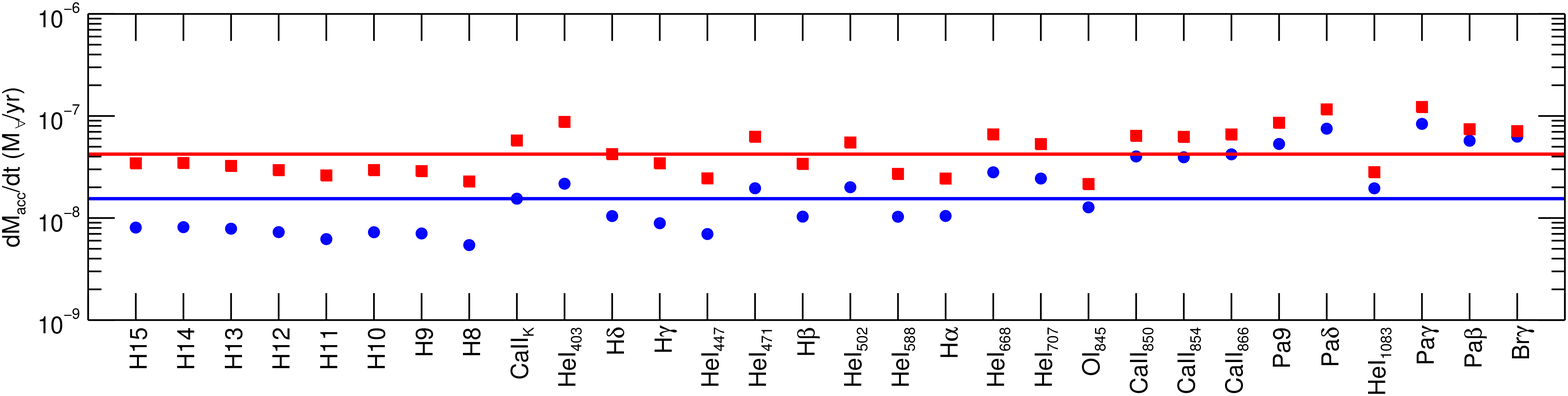}
\caption{Mass accretion rate determination computed from the flux of the labeled tracers using Alcal\'a et al. (2014) relationships. The assumed visual extinction is \av\,= 1 mag (blue circles) and \av\,=2 mag (red squares). The median \macc\, values are marked for both cases with a solid line.\label{fig:fig3}}
\end{figure}

An estimate of the mass accretion rate (\macc) was obtained from fluxes of bright permitted lines. We employed the relationships between line flux and accretion luminosity 
(\lacc) given by Alcal\'a et al. (2014). For each line, the \lacc\, value was then converted into mass accretion rate (\macc)  by adopting the equation by Gullbring et al. (1998). To have a meaningful comparison with the quiescence phase estimate, we have employed the same lines as in Paper I, namely \hi\, 
lines of the Balmer and Paschen series along with \hei\, and \caii\, bright lines. We also adopted the same parameters as in Paper I, namely: distance of 400 pc,
\mstar\,=\,0.4 \msun\, \rstar\,=\,1.29 \rsun\, (Hillenbrand 1997; Stassun et al. 1999), and inner radius \rin\,=\,5 \rsun. The result is depicted in Figure\,\ref{fig:fig3}, where we plot 
\macc\, for each of the considered lines. The median of the \macc\, values is 1.5 10$^{-8}$ \msunyr\, and 4.2  10$^{-8}$ \msunyr\, if the observed fluxes are de-reddened for A$_V$=\,1 and 2 mag, respectively. With the same extinction values, we get \macc\,=1$-$3 10$^{-9}$ \msunyr\, in the quiescence state (Paper I).

Assuming that the nearly tenfold enhancement of the \oi6300 flux is due to an increase of the mass ejection activity, we derived mass ejection 
rates  \mloss = 0.8 10$^{-9}$\,\msunyr\, and 2.0 10$^{-9}$\,\msunyr\ for A$_V$\,=\,1 and 2 mag, respectively, following the relationship given by Hartigan et al.
(1994). The same model gives 0.8 10$^{-10}$\,\msunyr\, and 
2.0 10$^{-10}$\,\msunyr\, for the quiescence phase. It is of interest to compare the ratio \macc/\mloss\, in the two phases; while this ratio is $\sim$ 13 in quiescence, it increases to $\sim$ 18 during the outburst. Although the difference is slight, this indicates that the enhancement of the mass ejection rate proceeds slower than that of mass accretion rate.

Finally, to compare the mass accretion rate in two subsequent outbursts, we applied the procedure described above to the fluxes reported in  Lorenzetti et al. (2006). 
We get  \macc\,(2005) $\sim$ 10$^{-7}$ \msunyr\, namely about a factor 5$-$10 larger than the present value. However, such estimate is based only on near-IR line fluxes, which, as can been seen in Figure\,\ref{fig:fig3}, tend to give higher values than the optical lines. In this sense, the value of \macc\,(2005)  has to be considered as an upper limit. Alternatively, this might indicate that the present outburst is weaker than the previous one or that the observed brightness is not at its maximum level. \\

\section{Concluding remarks}
We have detected a new outburst  of the EXor variable V1118 Ori, and obtained the optical and near-IR low-resolution spectra. These have been analyzed in comparison to the quiescence ones, which were acquired with a very similar instrumental setup. The main results of our study are:
\begin{itemize}
\item[-] we registered an increase of the continuum level. This increase goes from a factor $\sim$ 40 in $U$ band down to factor $\sim$ 4  $K$ band;
\item[-] the outburst spectrum is rich in emission lines, mainly from neutral and ionized metals. In particular, the fluxes of HI recombination lines have increased by about an order of magnitude;
\item[-] an increase of about an order of magnitude is estimated in the mass accretion rate. From bright optical and near-IR lines we estimate \macc\,=\, 1.2$-$4.8 10$^{-8}$\, \msunyr. 
\item[-] a few forbidden lines from OI and SII are detected, likely indicating an enhancement of the mass ejection activity that is associated with the increase of the accretion rate. From the \oi6300 luminosity we estimate \mloss\,=\, 0.8$-$2 10$^{-9}$\,\msunyr.
\item[-] a comparison with  light curves of previous outbursts suggests that the present outburst is relatively weak. 
Alternatively, it may be that the maximum level of the outburst activity was reached between January and September 2015 (when no observations are available), or that the outburst is still in the rising phase.
\end{itemize} 

\begin{table}
\footnotesize
\caption{\label{tab:tab1} Near-IR photometry.} 
\begin{center}
\begin{tabular}{ccccc}
\hline\\[-5pt]
Date          	& JD         & J	     &  H  		& K       \\
         	    & +2400000   &     \multicolumn{3}{c}{(mag)}  \\
\hline    
2014 Oct 19     & 56948.62   & 12.63     & 11.57    & 10.87   \\ 
2014 Oct 28     & 56958.63   & 12.62     & 11.56    & 10.79   \\
2014 Nov 01     & 56962.57   & 12.52     & 11.44    & 10.66   \\
2014 Nov 14     & 56975.55   & 12.64     & 11.61    & 10.93   \\
2014 Dec 23     & 57015.47   & 12.63     & 11.62    & 10.95   \\
2015 Jan 15     & 57038.38   & 12.63     & 11.58    & 10.87   \\
2015 Sep 22     & 57287.67   & 11.36     & 10.67    & 9.93    \\    
2015 Nov 09     & 57336.55   & 11.20     & 10.49    & 9.70    \\
2015 Nov 12     & 57339.57   & 11.38     & 10.59    & 9.79    \\
2015 Nov 13     & 57340.52   & 11.43     & 10.61    & 9.84    \\
2015 Nov 15     & 57342.54   & 11.64     & 10.84    & 10.04   \\
2015 Nov 17     & 57344.54   & 11.49     & 10.72    & 9.95    \\
2015 Nov 18     & 57345.54   & 11.46     &  -       &  -      \\
2015 Nov 19     & 57346.46   & 11.44     & 10.69    & 9.94    \\
2015 Dec 30     & 57357.45   & 11.32     & 10.60    & 9.84    \\
2015 Dec 01     & 57358.48   & 11.23     & 10.52    & 9.76    \\
2015 Dec 02     & 57359.43   & 11.42     & 10.64    & 9.83    \\
2015 Dec 02     & 57359.58   & 11.44     & 10.68    & 9.88    \\
2015 Dec 03     & 57360.47   & 11.37     & 10.65    & 9.87    \\
2015 Dec 04     & 57361.49   & 11.49     & 10.69    & 9.94    \\
2015 Dec 05     & 57362.41   & 11.61     & 10.79    & 10.01   \\
\hline
 \end{tabular} 
\end{center}
Note : Errors on near-IR magnitudes do not exceed 0.03 mag.
\end{table}

\begin{table}
\scriptsize
\caption{\label{tab:tab2} Main lines detected in V1118 Ori.} 
\begin{center}
\begin{tabular}{cccccc}
\hline\\[-5pt]
 Line ID	& $\lambda_{air}$   & Flux$\pm\Delta$(Flux)	             &  EW	  		& F$_{out}$/F$_{quiesc}$  & EW$_{out}$/EW$_{quiesc}$\\
		    &  (\AA)            & (10$^{-15}$ erg s$^{-1}$ cm$^{-2}$)&  (\AA)		&			              &                         \\
\hline 
\multicolumn{6}{c}{LBT-MODS}\\
\hline
  H15    	&  3711.97       	&   2.8$\pm$0.3		                 &  -1.4  		& 31$\pm$13               &   0.8                   \\
  H14    	&  3721.94	        &   3.1$\pm$0.3		                 &  -1.4  		& 34$\pm$14               &   0.9                   \\
  H13    	&  3734.37          &   4.3$\pm$0.4		                 &  -2.0  		& 48$\pm$20               &   1.4                   \\
  H12    	&  3750.15     	    &   4.8$\pm$0.4		                 &  -2.1 		& 30$\pm$8                &   0.9                   \\
  H11	    &  3770.63      	&   5.3$\pm$0.4	                     &  -2.5        & 16$\pm$3                &   0.5                   \\
  H10	    &  3797.90      	&   7.5$\pm$0.4	                     &  -3.5        & 15$\pm$2                &   0.5                   \\
  H9	    &  3835.38      	&   9.0$\pm$0.4	                     &  -3.9        & 10$\pm$1                &   0.4                   \\
  H8        &  3889.05          &   8.7$\pm$0.4                      &  -3.7        &  7$\pm$1                &   0.3                   \\
 CaII K     &  3933.66          &  41.2$\pm$0.4                      &  -18.0       & 91$\pm$5                &   3.0                   \\
 CaII+H7    &  3968.45/3970.07  &  40.0$\pm$0.4                      &  -17.4       & 17$\pm$1                &   0.7                   \\
 HeI        &  4026.2           &   3.2$\pm$0.4                      &  -1.5        & 18$\pm$5                &   0.9                   \\
 H$\delta$  &  4101.73          &  22.5$\pm$0.4                      &  -9.6        & 8$\pm$1    	          &   0.4                   \\
 H$\gamma$  &  4340.46          &  26.0$\pm$0.4                      &  -9.1        & 6.0$\pm$0.2             &   0.3                   \\
 HeI        &  4471.5           &   2.0$\pm$0.4                      &  -0.7        & 7$\pm$2                 &   0.2                   \\
 HeI        &  4713.2           &   0.9$\pm$0.4                      &  -0.3        & 10$\pm$6                &   0.6                   \\
 H$\beta$   &  4861.32          &  59.9$\pm$0.4                      &  -18.5       & 9.0$\pm$0.2             &   0.5                   \\
 HeI        &  5015.67          &   4.1$\pm$0.4                      &  -1.2        & 24$\pm$5                &   1.3                   \\
 HeI        &  5875.6           &   7.3$\pm$0.4                      &  -2.3        & 7$\pm$1                 &   0.4                   \\
 $[\rm{OI}]$&  6300.30          &   3.4$\pm$0.4                      &  -0.8        & 11$\pm$3                &   0.8                   \\
 H$\alpha$  &  6562.80          & 470.3$\pm$0.4                      &  -94.6       & 8.0$\pm$0.1             &   0.8                   \\
 HeI        &  6678.15          &   7.4$\pm$0.4                      &  -1.5        & 8$\pm$1                 &   0.7                   \\
$[\rm{SII}]$&  6716.44          &   1.3$\pm$0.4                      &  -0.3        & 4$\pm$2                 &   0.5                   \\
$[\rm{SII}]$&  6730.81          &   $<$1.2                           &   -          & $<$3.5                  &   -                     \\
 HeI        &  7065.2           &   5.4$\pm$0.4                      &  -1.0        & 8$\pm$3                 &   0.7                   \\
 OI         &  8446.5           &  18.3$\pm$0.4                      &  -3.2        & 12$\pm$2                &   2.0                   \\
 CaII       &  8498.03          & 187.6$\pm$0.4                      &  -33.1       & 83$\pm$4                &  14.4                   \\
 CaII       &  8542.09          & 208.6$\pm$0.4                      &  -36.6       & 84$\pm$2                &  15.9                   \\
 CaII       &  8662.14          & 191.1$\pm$0.4                      &  -34.8       & 92$\pm$4                &  19.3                   \\
 Pa12       &  8750.47          &  33.8$\pm$0.4                      &  -6.1        & 28$\pm$5                &   6.7                   \\
 Pa9        &  9229.01          &  60.1$\pm$0.4                      & -11.7        & 25$\pm$9                &   7.9                   \\
\hline
\multicolumn{6}{c}{LBT-LUCI2}\\
\hline
Pa$\delta$ & 10049.37          & 135$\pm$10                          & -14          &   16$\pm$3              &   4.7                   \\ 
HeI        & 10830.3           & 118$\pm$10                          & -11          &   17$\pm$3              &   4.5                   \\ 
Pa$\gamma$ & 10938.09          & 212$\pm$10                          & -19          &   20$\pm$3              &   4.4                   \\ 
Pa$\beta$  & 12818.08          & 323$\pm$10                          & -29          &   12$\pm$32             &   2.1                   \\ 
MgI  	   & 15029/15051       &  71$\pm$7                           &  -8          &   $>$11                 &    -                    \\
Br19       & 15264.71          &  30$\pm$7                           &  -4          &   $>$6                  &     -                   \\ 
Br18       & 15345.98          &  32$\pm$7                           &  -4          &   $>$6                  &     -                   \\
Br17       & 15443.14          &  35$\pm$7                           &  -5          &   $>$7                  &     -                   \\
Br16       & 15560.70          &  37$\pm$7                           &  -6          &   $>$6                  &     -                   \\  
Br15       & 15704.95          &  41$\pm$7                           &  -6          &   $>$8                  &     -                   \\
Br14+OI    & 15885/15892       &  86$\pm$7                           & -12          &   $>$14                 &     -                   \\ 
Br13       & 16113.71          &  54$\pm$7                           &  -8          &   $>$9                  &     -                   \\
Br12       & 16411.67          &  62$\pm$7                           &  -9          &   $>$10                 &     -                   \\
Br11       & 16811.11          &  72$\pm$7                           & -11          &   $>$12                 &     -                   \\
Br10       & 17366.85          &  75$\pm$7                           & -11          &   $>$12                 &     -                   \\
Br$\gamma$ & 21655.29          &  78$\pm$6                           & -16          &   $>$13                 &     -                   \\   
NaI        & 22062.4           &  7$\pm$1                            & -1.5         &   $>$1.2                &     -                   \\
NaI        & 22089.7           &  9$\pm$1                            & -2           &   $>$1.5                &     -                   \\
CO 2-0     & 22943             &  158$\pm$1                          & -34          &   $>$26                 &     -                   \\
CO 3-1     & 23235             &  142$\pm$1                          & -30          &   $>$24                 &     -                   \\     
\hline
 \end{tabular} 
\end{center}
Note : Upper limits are computed at 3$\sigma$ level.
\end{table}

\section{Acknowledgements}

Based on observations made with different instruments:[1] the Large Binocular Telescope (LBT). The LBT is an international collaboration among institutions in the United States, Italy and Germany. LBT Corporation partners are: The University of Arizona on behalf of the Arizona university system; Istituto Nazionale di Astrofisica, Italy; LBT Beteiligungsgesellschaft, Germany, representing the Max-Planck Society, the Astrophysical Institute Potsdam, and Heidelberg University; The Ohio State University, and The Research Corporation, on behalf of The University of Notre Dame, University of Minnesota and University of Virginia. [2] the AZT-24 IR Telescope at Campo Imperatore (L'Aquila - Italy) operated under the responsability of the INAF-Osservatorio Astronomico di Roma (OAR). V.L. acknowledges support from St.Petersburg University research grant 6.38.335.2015.

{}





\newpage
\normalsize

\begin{thebibliography}{}
\bibitem[Alcal{\'a} et al.(2014)]{2014A&A...561A...2A} Alcal{\'a}, J.~M., Natta, A., Manara, C.~F., et al.\ 2014, \aap, 561, 2 
\bibitem[Antoniucci et al.(2014)]{2014A&A...565L...7A} Antoniucci, S., Arkharov, A.~A., Di Paola, A., et al.\ 2014, \aap, 565, L7 
\bibitem[Antoniucci et al.(2008)]{2008A&A...479..503A} Antoniucci, S., Nisini, B., Giannini, T., \& Lorenzetti, D.\ 2008, \aap, 479, 503 
\bibitem{} Antoniucci, S., Rigliaco, E., Nisini, B., et al. 2016, submitted
\bibitem[Audard et al.(2014)]{2014prpl.conf..387A} Audard, M., {\'A}brah{\'a}m, P., Dunham, M.~M., et al.\ 2014, Protostars and Planets VI, 387 
\bibitem[Audard et al.(2005)]{2005ApJ...635L..81A} Audard, M., G{\"u}del, M., Skinner, S.~L., et al.\ 2005, \apjl, 635, L81 
\bibitem[Audard et al.(2016)]{2016ATel.8548....1A} Audard, M., Hamaguchi, K., Kastner, J., Grosso, N., \& Walter, F.~M.\ 2016, The Astronomer's Telegram, 8548
\bibitem[Audard et al.(2010)]{2010A&A...511A..63A} Audard, M., Stringfellow, G.~S., G{\"u}del, M., et al.\ 2010, \aap, 511, 63 
\bibitem[Black \& van Dishoeck(1987)]{1987ApJ...322..412B} Black, J.~H., \& van Dishoeck, E.~F.\ 1987, \apj, 322, 412 
\bibitem[D'Alessio et al.(2000)]{2000SPIE.4008..748D} D'Alessio, F., Di Cianno, A., Di Paola, A., et al.\ 2000, \procspie, 4008, 748 
\bibitem[Edwards et al.(2013)]{2013ApJ...778..148E} Edwards, S., Kwan, J., Fischer, W., et al.\ 2013, \apj, 778, 148 
\bibitem[Garcia Garcia et al.(1995)]{1995IBVS.4268....1G} Garcia Garcia, J., Mampaso, A., \& Parsamian, E.~S.\ 1995, Information Bulletin on Variable Stars, 4268, 1 
\bibitem[Garcia Garcia \& Parsamian(2000)]{2000IBVS.4925....1G} Garcia Garcia, J., \& Parsamian, E.~S.\ 2000, Information Bulletin on Variable Stars, 4925, 1 
\bibitem[Gasparian \& Ohanian(1989)]{1989IBVS.3327....1G} Gasparian, K.~G., \& Ohanian, G.~B.\ 1989, Information Bulletin on Variable Stars, 3327, 1 
\bibitem[Gullbring et al.(1998)]{1998ApJ...492..323G} Gullbring, E., Hartmann, L., Brice{\~n}o, C., \& Calvet, N.\ 1998, \apj, 492, 323 
\bibitem[Hartigan et al.(1994)]{1994ApJ...436..125H} Hartigan, P., Morse, J.~A., \& Raymond, J.\ 1994, \apj, 436, 125 
\bibitem[Hartmann \& Kenyon(1985)]{1985ApJ...299..462H} Hartmann, L., \& Kenyon, S.~J.\ 1985, \apj, 299, 462 
\bibitem[Hayakawa et al.(1998)]{1998IBVS.4615....1H} Hayakawa, T., Ueda, T., Uemura, M., et al.\ 1998, Information Bulletin on Variable Stars, 4615, 1 
\bibitem[Herbig(1989)]{1989ESOC...33..233H} Herbig, G.~H.\ 1989, European Southern Observatory Conference and Workshop Proceedings, 33, 233 
\bibitem[Herbig(2008)]{2008AJ....135..637H} Herbig, G.~H.\ 2008, \aj, 135, 637 
\bibitem[Herczeg \& Hillenbrand(2008)]{2008ApJ...681..594H} Herczeg, G.~J., \& Hillenbrand, L.~A.\ 2008, \apj, 681, 594 
\bibitem[Hillenbrand(1997)]{1997AJ....113.1733H} Hillenbrand, L.~A.\ 1997, \aj, 113, 1733 
\bibitem[K{\'o}sp{\'a}l et al.(2011)]{2011A&A...527A.133K} K{\'o}sp{\'a}l, {\'A}., {\'A}brah{\'a}m, P., Acosta-Pulido, J.~A., et al.\ 2011, \aap, 527, A133 
\bibitem[Lorenzetti et al.(2012)]{2012ApJ...749..188L} Lorenzetti, D., Antoniucci, S., Giannini, T., et al.\ 2012, \apj, 749, 188 
\bibitem[Lorenzetti et al.(2015)]{2015ApJ...802...24L} Lorenzetti, D., Antoniucci, S., Giannini, T., et al.\ 2015a, \apj, 802, 24 (Paper I)
\bibitem[Lorenzetti et al.(2015)]{2015ATel.8100....1L} Lorenzetti, D., Arkharov, A.~A., Di Paola, A., et al.\ 2015b, The Astronomer's Telegram, 8100
\bibitem[Lorenzetti et al.(2006)]{2006A&A...453..579L} Lorenzetti, D., Giannini, T., Calzoletti, L., et al.\ 2006, \aap, 453, 579 
\bibitem[Lorenzetti et al.(2007)]{2007ApJ...665.1182L} Lorenzetti, D., Giannini, T., Larionov, V.~M., et al.\ 2007, \apj, 665, 1182 
\bibitem[Lorenzetti et al.(2009)]{2009ApJ...693.1056L} Lorenzetti, D., Larionov, V.~M., Giannini, T., et al.\ 2009, \apj, 693, 1056 
\bibitem[Manara et al.(2016)]{2016A&A...585A.136M} Manara, C.~F., Fedele, D., Herczeg, G.~J., \& Teixeira, P.~S.\ 2016, \aap, 585, 136 
\bibitem[Marsden(1984)]{1984IAUC.3959....2M} Marsden, B.~G.\ 1984, \iaucirc, 3924 
\bibitem[Parsamyan et al.(1993)]{1993Ap.....36...12P} Parsamian, E.~S., Ibragimov, M.~A., Oganyan, G.~B., \& Gasparyan, K.~G.\ 1993, Astrophysics, 36, 12 
\bibitem[Parsamian et al.(2002)]{2002Ap.....45..393P} Parsamian, E.~S., Mujica, R., \& Corral, L.\ 2002, Astrophysics, 45, 393 
\bibitem[Pogge et al.(2010)]{2010SPIE.7735E..0AP} Pogge, R.~W., Atwood, B., Brewer, D.~F., et al.\ 2010, \procspie, 7735, 77350A 
\bibitem[Scoville et al.(1980)]{1980ApJ...240..929S} Scoville, N.~Z., Krotkov, R., \& Wang, D.\ 1980, \apj, 240, 929 
\bibitem[Shu et al.(1994)]{1994ApJ...429..797S} Shu, F.~H., Najita, J., Ruden, S.~P., \& Lizano, S.\ 1994, \apj, 429, 797 
\bibitem[Sicilia-Aguilar et al.(2012)]{2012A&A...544A..93S} Sicilia-Aguilar, A., K{\'o}sp{\'a}l, {\'A}., Setiawan, J., et al.\ 2012, \aap, 544, 93 
\bibitem[Stassun et al.(1999)]{1999AJ....117.2941S} Stassun, K.~G., Mathieu, R.~D., Mazeh, T., \& Vrba, F.~J.\ 1999, \aj, 117, 2941 
%
\end{thebibliography}
\end{document}